\documentclass[aps,prx,reprint,showpacs,superscriptaddress,floatfix]{revtex4-2}
\usepackage{amsmath,amssymb,amsfonts,mathtools}
\usepackage{graphicx}
\usepackage{xcolor}
\usepackage[colorlinks=true,allcolors=blue]{hyperref}
\usepackage[T1]{fontenc}
\usepackage{lmodern}
\usepackage{amsmath,amssymb,amsfonts,mathtools}
\usepackage{bm}
\usepackage[normalem]{ulem}
\usepackage{graphicx}
\usepackage{xcolor}
\usepackage{booktabs}
\usepackage{multirow}
\usepackage{slashed}
\usepackage{ragged2e}
\usepackage{comment}
\usepackage{xspace}
\usepackage{mathrsfs}
\usepackage{orcidlink}
\usepackage{listings}
\usepackage{tikz}
\usetikzlibrary{arrows,arrows.meta}
\usepackage[braket,qm]{qcircuit}
\usepackage[normalem]{ulem}

\makeatletter
\renewcommand{\onecolumngrid}{%
  \do@columngrid{one}{\@ne}%
  \let\set@footnotewidth\set@footnotewidth@one
  \let\compose@footnotes\compose@footnotes@one
}
\makeatother

\newcommand{\jmax}{j_{\mathrm{max}}}
\newcommand{\jcut}{j_{\text{cut}}}

\newcommand{\HA}{\mathcal{H}_A}
\newcommand{\rhoA}{\rho_A}

\definecolor{forestgreen}{rgb}{0.13, 0.55, 0.13}

\definecolor{green}{RGB}{34,139,34}

\newcommand{\Tr}{\operatorname{Tr}}

\newcommand{\hAone}{\ensuremath{\hat{A}^{(1)}}}
\newcommand{\hAtwo}{\ensuremath{\hat{A}^{(2)}}}
\newcommand{\hBone}{\ensuremath{\hat{B}^{(1)}}}
\newcommand{\hBtwo}{\ensuremath{\hat{B}^{(2)}}}

\newcommand{\half}{\tfrac{1}{2}}
\newcommand{\jL}{j_L}
\newcommand{\jR}{j_R}
\newcommand{\jM}{j_M}
\newcommand{\nM}{n_M}

\newcommand{\dphys}{d_{\mathrm{phys}}}
\newcommand{\sixj}[6]{\begin{Bmatrix}#1&#2&#3\\#4&#5&#6\end{Bmatrix}}

\begin{document}

\title{Magic and entanglement in 1+1-dimensional SU(2) lattice gauge theory}

\author{Raghav~G.~Jha\orcidlink{0000-0003-2933-0102}}
\email{raghav.govind.jha@gmail.com}
\affiliation{Department of Physics and Astronomy, North Carolina State University, Raleigh, North Carolina 27695, USA}

\author{Goksu~C.~Toga\orcidlink{0000-0002-0316-2502}}
\email{gctoga@ncsu.edu}
\affiliation{Department of Physics and Astronomy, North Carolina State University, Raleigh, North Carolina 27695, USA}

\author{Jaber~I.~Taher\orcidlink{0009-0003-0189-4317}}
\email{mtaher@ncsu.edu}
\affiliation{Department of Physics and Astronomy, North Carolina State University, Raleigh, North Carolina 27695, USA}

\author{Bojko~N.~Bakalov\orcidlink{0000-0003-4630-6120}}
\email{bnbakalo@ncsu.edu}
\affiliation{Department of Mathematics, North Carolina State University, Raleigh, North Carolina 27695, USA}

\author{Alexander~F.~Kemper\orcidlink{0000-0002-5426-5181}}
\email{akemper@ncsu.edu}
\affiliation{Department of Physics and Astronomy, North Carolina State University, Raleigh, North Carolina 27695, USA}

\begin{abstract}
Entanglement and non-stabilizerness (magic) quantify two distinct departures of quantum systems from classical description: the former measures non-local correlations, while the latter measures the deviation from stabilizer states that can be efficiently simulated classically. Understanding magic in physically relevant quantum field theories is essential for identifying where quantum advantage may be realized in the early fault-tolerant quantum computing era. 
We calculate the gauge-invariant entanglement entropy and stabilizer R\'{e}nyi entropy of the ground state of the (1+1)-dimensional SU(2) lattice gauge theory formulated in a dressed-site basis that enforces Gauss's law exactly. Using tensor networks, we obtain results for system sizes up to $L=100$ (300 qubits).  We find a crossover denoted by $g_{\star}$ where the ground state passes from a more magic-rich regime into a regime with less magic;
this is also tracked by the sharpest change of both the entanglement entropy and lattice particle density. Our large-scale study of non-stabilizerness and entanglement entropy in a non-Abelian lattice gauge theory with matter provides new insight into the interplay of magic and entanglement in gauge theories, both of which are relevant for classical and early fault-tolerant quantum simulations.

\end{abstract}

\maketitle

\textit{Introduction} --- 
Lattice gauge theories (LGTs) are expected to be one of the main applications of future quantum computers~\cite{Feynman:1981tf, Preskill:2018fag},
particularly non-Abelian theories where the fermionic sign problem and real-time dynamics remain largely intractable classically. A central question
is: what does quantum information theory tell us about the classical and quantum simulations of LGTs? This question is crucial for quantum simulations of LGTs in the coming years as we move away from noisy intermediate-scale quantum simulations towards early fault-tolerant quantum simulations. 

Entanglement has long served as the primary diagnostic of quantum correlations in gauge theories. However, entanglement alone does not determine classical simulability. The Gottesman-Knill theorem~\cite{Gottesman:1998hu} establishes that Clifford circuits with stabilizer inputs and Pauli measurements can be efficiently simulated classically, regardless of the entanglement they
generate~\cite{Aaronson:2004xuh}. It is the addition of non-stabilizer
resources; quantified by measures of \emph{non-stabilizerness} or
``magic''~\cite{Bravyi:2004isx} that renders a quantum computation
classically intractable. A highly entangled stabilizer state is classically easy; a product state doped with magic is not. Magic and entanglement thus capture fundamentally different aspects of quantum complexity: entanglement measures nonlocal correlations, while magic measures the distance from the efficiently simulable set of stabilizer states.

\begin{figure}[t]
    \centering
    \includegraphics[width=0.40\textwidth]{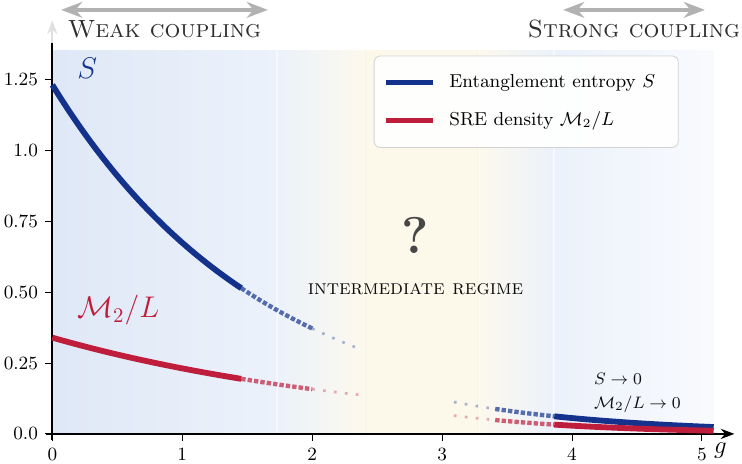}
    \caption{Schematic representation of gauge-invariant entanglement entropy and stabilizer R\'{e}nyi entropy (SRE) as a function of the coupling parameter ($g$) for some fixed value of $m$. In this paper, we find a crossover coupling $g_{\star}$ where these observables or their derivatives have a major change in behavior.}
    \label{fig:fig1_plan}
\end{figure}

This distinction has motivated a rapidly growing body of work on non-stabilizerness in quantum many-body systems~\cite{Haug:2022vpg, Tarabunga:2023ggd, Niroula:2023meg, Tarabunga:2024ugl}. Within the context of LGTs, non-stabilizerness has only recently been studied, in contexts such as Abelian $\mathrm{U}(1)$ theory~\cite{Falcao:2024msg}, string breaking ~\cite{Grieninger:2026bdq}, and in discrete and non-Abelian \emph{pure-gauge} theories on a ladder~\cite{Esposito:2024uzw, Santra:2025dsm}. 
For non-Abelian gauge theories \emph{with matter}, however, no study of non-stabilizerness exists to date, to our knowledge. In this work, we study the $(1+1)$-dimensional $\mathrm{SU}(2)$ lattice gauge theory with staggered fermions at the hardcore-gluon ($j_{\max} = 1/2$) truncation, using a gauge-invariant dressed-site basis that exactly enforces Gauss's law.
While we recover expected results in the extreme ($g = 0, g \to \infty$)  limits, in the intermediate regime, we find a clear \emph{decoupling} of the two measures around $g_{\star}$, where the entanglement entropy decreases monotonically with increasing coupling $g$, while the SRE exhibits non-monotonic behavior, retaining significant non-stabilizerness in a regime where entanglement has already been substantially suppressed.
We show the schematic representation of the goal of the paper in Fig.~\ref{fig:fig1_plan}.  

\begin{figure}
    \centering
    \includegraphics[width=0.48\textwidth]{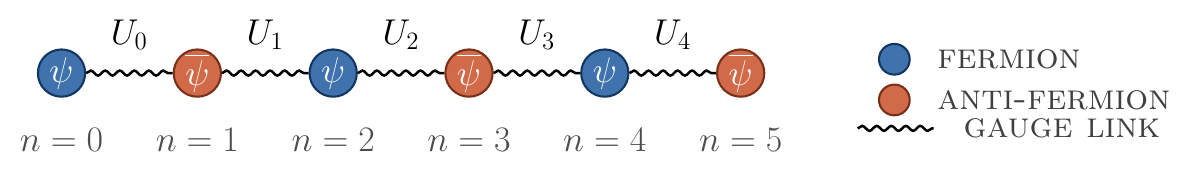}
    \par\vspace{0.2cm}
    \includegraphics[width=0.48\textwidth]{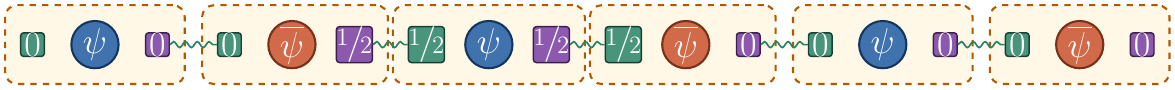}
    \caption{Spatial lattice with staggered fermions and gauge links $U_n$ pointing from site $n$ to $n+1$ (top) and example of a gauge-singlet state (bottom).}
    \label{fig:cartoon_dressedSU2}
\end{figure}

%%%%%%%%%%%%%%%%%%%%%%%%%%%%

\textit{Dressed-site formulation of $\mathrm{SU}(2)$ lattice gauge theory} --- The $(1+1)$-dimensional $\mathrm{SU}(2)$ lattice gauge theory with staggered fermions is governed by the Hamiltonian~\cite{Kogut:1974ag}:
\begin{align}
H &= g^2\sum_{n=1}^{L-1} \sum_{A=1}^{3} \big(E_{n,n+1}^{A}\big)^2 + m \sum_{n=1}^{L}(-1)^n \sum_{a=\textcolor{red}{r},\,\textcolor{forestgreen}{g}}
\psi_{n,a}^\dagger \psi_{n,a} \nonumber \\
&\quad + \sum_{n=1}^{L-1}\sum_{a,b=\textcolor{red}{r},\,\textcolor{forestgreen}{g}}
\left(
\psi_{n,a}^\dagger\, U_{n,n+1}^{ab}\, \psi_{n+1,b}
+\mathrm{h.c.}
\right),
\label{eq:Hamiltonian_standard}
\end{align}
where $E^{A}_{n,n+1}$ are the three $\mathrm{SU}(2)$ color--electric field operators on the link connecting sites $n$ and $n+1$, $U_{n,n+1}$ is the parallel-transport link operator, and $\psi_{n,a}$ are single-component staggered fermion fields with color index $a\in\{\textcolor{red}{r},\textcolor{forestgreen}{g}\}$. In Eq.~\eqref{eq:Hamiltonian_standard}, $m$ and $g$ are dimensionless parameters that are related to dimensionful parameters through appropriate powers of lattice spacing ($a$) i.e., 
$g = ag_{\text{dim}}$ and $m = am_{\text{physical}}$. Since the Hamiltonian is one-dimensional, no magnetic (plaquette) term appears.
Out of the total possible states, the physical states are the ones 
annihilated by the generators of Gauss's law,
$G^{A}_n\ket{\psi} = 0$ for all sites $n$, where $G^{A=1,2,3}_n$ are the three local gauge generators. This requires combining three quantum numbers i.e., two from the each half-links and one from the matter sector, such that they couple to a color singlet. We provide more details in Appendix~\ref{app:su2_rep_gi_part}.

In the dressed-site formulation, the matter field at site $n$ and its adjacent half-links are combined into a single composite object as shown in Fig.~\ref{fig:cartoon_dressedSU2}.  
Before imposing gauge invariance, the microscopic Hilbert space for $\jmax=1/2$ is
$(\mathbf{1}\oplus\mathbf{2}) \otimes (\mathbf{1}\oplus\mathbf{2}\oplus\mathbf{1}) \otimes (\mathbf{1}\oplus\mathbf{2}) = 36$
states per site. The local states can be labeled by the quantum numbers $\ket{\jL,\nM,\jR}$, where $\nM\in\{0,1,2\}$ is the matter occupation number. For the irreducible $\mathrm{SU}(2)$-representation (hereafter referred to as `irrep') with cutoff $\jmax=1/2$, the allowed gauge representations are $j = 0$ and $j = \tfrac{1}{2}$. At a staggered matter site with one fermion flavor and two colors, the local matter Fock space is
\begin{equation}
\mathcal{H}_M = \mathrm{span}\!\left\{
\ket{0},\;
(\psi_{n}^{\textcolor{red}{r}})^\dagger\ket{0},\;
(\psi_{n}^{\textcolor{forestgreen}{g}})^\dagger\ket{0},\;
(\psi_{n}^{\textcolor{red}{r}})^\dagger(\psi_{n}^{\textcolor{forestgreen}{g}})^\dagger\ket{0}
\right\},
\end{equation}
which decomposes under $\mathrm{SU}(2)$ color as $\mathcal{H}_M \simeq \mathbf{1} \oplus \mathbf{2} \oplus \mathbf{1}$. The vacuum $\ket{0}\equiv\ket{j{=}0,\,m{=}0}$ and the doubly occupied state
$\ket{d} = (\psi_n^{\textcolor{red}{r}})^\dagger(\psi_n^{\textcolor{forestgreen}{g}})^\dagger\ket{0}$
are both color singlets ($\jM = 0$), while the single-quark states
$\ket{\textcolor{red}{r}} \equiv (\psi_n^{\textcolor{red}{r}})^\dagger\ket{0}$
and $\ket{\textcolor{forestgreen}{g}} \equiv (\psi_n^{\textcolor{forestgreen}{g}})^\dagger\ket{0}$
form a color doublet ($\jM = 1/2$).
In the dressed-site basis, Gauss's law is enforced locally by requiring that the left flux $\jL$, the matter representation $\jM$, and the right flux $\jR$ couple to an overall $\mathrm{SU}(2)$ singlet,
i.e., that $\jL \otimes \jM \otimes \jR$ fuses to total spin zero. This reduces the number of states down from 36 to 6. Further details are provided in Appendix~\ref{app:su2_rep_gi_part}.

In this physical basis, the Hamiltonian of Eq.~\eqref{eq:Hamiltonian_standard} takes the form~\cite{Calajo:2024qrc}:
\begin{equation}
  H = \sum_n \Big[g^2 \hat{C}_n + m\,(-1)^n \hat{M}_n + \hAone_{n}\hBone_{n+1} + \hAtwo_{n}\hBtwo_{n+1}\Big],
  \label{eq:Hamiltonian_dressed}
\end{equation}
where $\hat{M}_n$ and $\hat{C}_n$ are diagonal $6 \times 6$ operators encoding the mass and electric-field contributions respectively, $(-1)^n$ is the staggered phase factor and $\hAone$, $\hBone$, $\hAtwo$, $\hBtwo$ are hopping operators acting within the physical subspace. We provide details of the construction in Appendix~\ref{app:sec_H_cons}.

\begin{figure}
    \centering
    \includegraphics[width=1.0\linewidth]{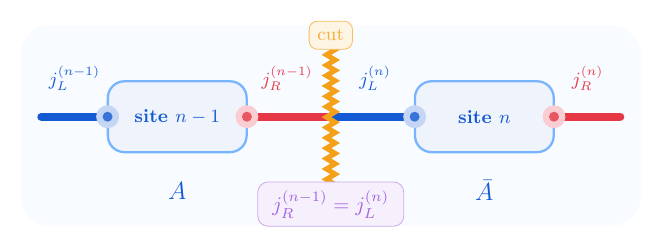}
    \caption{Schematic representation of a two-site system in the dressed-site basis. Each dressed site carries a matter occupation number $n_M$ and left/right half-link irreps $j_L$ and $j_R$.
    The physical link connecting sites $n{-}1$ and $n$ is shared by
    $j_R^{(n-1)}$ and $j_L^{(n)}$, which satisfy the link constraint
    $j_R^{(n-1)}=j_L^{(n)}$. A bipartition through this link defines the gauge-invariant entanglement entropy, resolved by the irrep on the cut link.}
    \label{fig:EE_cut_fig}
\end{figure}

%%%%%%%%%%%%%%%%%%%%%

\textit{Entropy and magic at the CFT point $m=g=0$} --- To uncover the resource landscape of non-Abelian gauge theories and probe the interplay between different quantum resources relevant to classical/quantum simulations, we investigate three observables across the parameter space: gauge-invariant entanglement entropy, stabilizer R\'{e}nyi entropy, and the particle density. 

We first start with a particular limit of the Hamiltonian given in Eq.~\eqref{eq:Hamiltonian_dressed} by setting $m = g = 0$.  
This pure-hopping model flows to a low-energy theory which is a $c=1$ conformal field theory (CFT). 
For open boundary conditions (OBC), the well-known result for the bipartite entanglement entropy is~\cite{Calabrese:2009qy, Bazavov:2017hzi}:
\begin{equation}
    S(l) = \frac{c}{6} \ln\Bigg[\frac{2L}{\pi a } \sin \Bigg(\frac{\pi l}{L}\Bigg)\Bigg] + \text{const},
\end{equation}
where $a$ is the lattice spacing or the UV cutoff, which we have set to $a=1$, and $l$ determines the location of the cut. 

To verify this behavior, we calculate the gauge-invariant entanglement entropy \cite{Donnelly:2011hn} using the setup shown in Fig.~\ref{fig:EE_cut_fig} at the central bond, i.e., $l = L/2$.  For a bipartition through the link connecting sites $n{-}1$ and $n$ indicated by the orange cut in Fig.~\ref{fig:EE_cut_fig}, the entanglement entropy decomposes into three physically distinct contributions~\cite{Donnelly:2011hn}:
\begin{equation}
    S = \underbrace{-\sum_j p_j \ln p_j}_{S_{\text{classical}}} + \underbrace{\sum_j p_j\, S_j}_{S_{\text{quantum}}} + \underbrace{\sum_j p_j \ln(d_j)}_{S_{\text{edge}}},
    \label{eq:EE_decomposition}
\end{equation}
where the sum runs over all boundary irreps $j \in \{0,\,1/2\}$ populated by the ground state at the cut. Here, $p_j$ is the probability of finding irrep $j$ on the cut link, $S_j$ is the entanglement entropy within the $j$-sector, and $d_j$ is the dimension of the irrep. For a generic gauge-invariant ground state, the wavefunction is a superposition of basis states carrying different flux values through the cut, so $S$ receives contributions from multiple irrep sectors weighted by $p_j$. We provide additional details about gauge-invariant entanglement entropy in Appendix~\ref{app:EE_SRE}.

\begin{figure}
    \centering
    \includegraphics[width=1.0\linewidth]{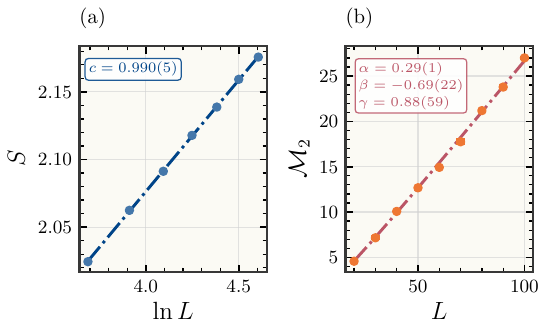}
    \caption{a) The gauge-invariant entanglement entropy at the central bond ($L/2$) at the CFT point for different $L$, obtaining $c = 0.990(5)$. b) SRE with $m = g = 0$ for different $L$ at the CFT point and obtain
    $\alpha = 0.29(1)$, $\beta = -0.69(22)$, and $\gamma = 0.88(59)$ by fitting to Eq.~\eqref{eq:M2-CFT-pred}.}
    \label{fig:result_CFT}
\end{figure}

We scale the system size and perform a linear fit to obtain the central charge as $c = 0.990(5)$, which is in agreement with the expected CFT results, as shown in Fig.~\ref{fig:result_CFT}. Even with a severe truncation $\jmax =  1/2$, we obtain an accurate estimate of the central charge. This indicates that the hardcore gluon truncation admits the same universality class as the full Hamiltonian lattice gauge theory, however, a systematic
comparison with larger $\jmax$ would be needed to establish this conclusively. 

 \begin{figure}
    \centering
\includegraphics[width=0.99\linewidth]{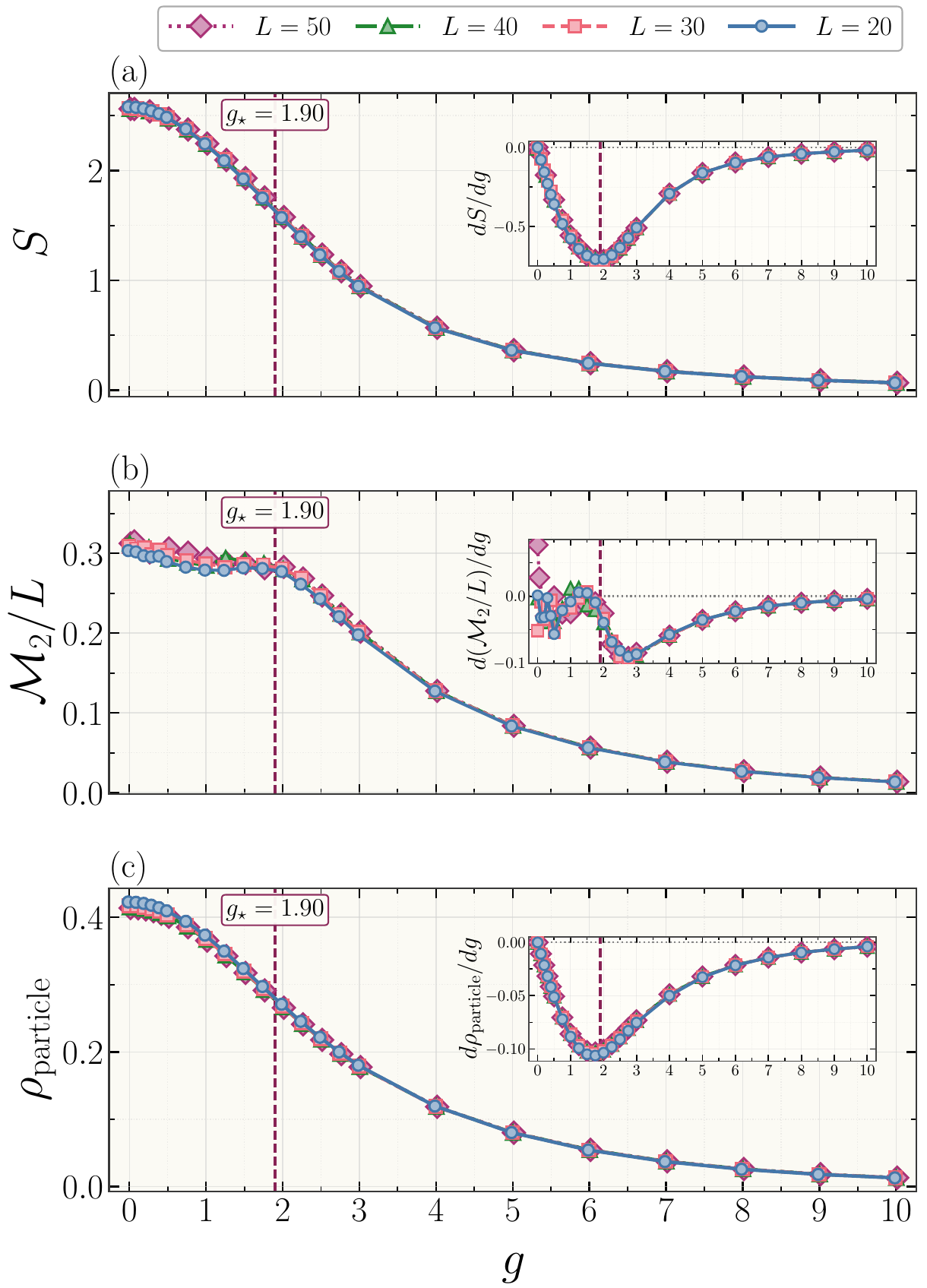}
    \caption{The entanglement entropy (a), volume-normalized SRE ($\mathcal{M}_2/L$) (b), and particle number density $(\rho_\text{particle})$ (c) for different $g$ for fixed $m=0.2$ for different lattice sizes. The inset shows the corresponding $g$ derivative of these observables. We find a crossover coupling, $g_{\star} \approx 1.9$. The results are obtained with a fixed maximum bond dimension of $\chi=100$.}
    \label{fig:six_panels}
\end{figure}

We next investigate non-stabilizerness, remaining still at the CFT limit, $m = g = 0$. We probe the magic in our model by calculating the stabilizer R\'{e}nyi-$n$ entropy $\mathcal{M}_{n}$: 
\begin{align}
\mathcal{M}_{n}
&= \frac{1}{1-n} \log_2 \Bigg(\frac{1}{2^{N}}
\sum_{P\in\mathcal{P}_N}
\bigl|\langle P\rangle\bigr|^{2n}\Bigg),
\label{eq:sre-alpha_per_site}
\end{align}
where $\mathcal{P}_N=\{I,X,Y,Z\}^{\otimes N}$ denotes the $N$-qubit Pauli group (excluding phases), $\langle P\rangle := \bra{\psi}P\ket{\psi}$ with $P\in\mathcal{P}_N$ is the expectation value of a Pauli string in a pure state $\ket{\psi} $, $N$ is the number of qubits, and $n$ is the R\'{e}nyi index. In this work, we use $n=2$, the most commonly employed order for characterizing magic~\cite{Leone:2021oqe}. Since SRE is an extensive quantity, we primarily report the normalized value $\mathcal{M}_2/L$. As is well-known, the ground states of CFT are magical~\cite{White2021CFTMagical}, and for our case, the maximum value for both EE and SRE is found to be at $m=g=0$. In other words, the gapless critical state (ground state of the CFT) maximizes quantum complexity or non-stabilizerness.
Specifically, the amount of magic follows a certain parametric dependence on $L$ given by~\cite{White2021CFTMagical,  Hoshino:2025ine, Trino:2025luw}:
\begin{equation}
    \mathcal{M}_{2} = \alpha L + \beta \ln(L) + \gamma, 
    \label{eq:M2-CFT-pred}
\end{equation}
and doing a fit to the data up to $L=100$, we find $\alpha = 0.29(1)$, $\beta= -0.69(22)$, and $\gamma = 0.88(59)$. In this expression, bulk magic density is captured by $\alpha$, which is non-universal and depends on lattice regularization and other microscopic details. The exponent, $\gamma$, is the constant piece which encodes $O(1)$ contributions. For the other analytically tractable limit, $g \to \infty$, the ground state is a product pure stabilizer state and has zero entropy and magic.

\textit{Moving toward confinement ---}
Next, we go away from the $m, g = 0$ limit to explore general $m-g$ parameter space, finding \emph{striking difference} in the behavior of entanglement and magic. We first consider a fixed value of $m = 0.2$ and show the dependence of the entanglement entropy for various system sizes $L$ in Fig.~\ref{fig:six_panels} (a).  
The theory is gapped, and the logarithmic dependence of entanglement entropy on system size is absent compared to the CFT. We also compute $\mathcal{M}_2/L$ for various $L$, and the results are shown in Fig.~\ref{fig:six_panels} (b). We observe a striking difference in the magic $\mathcal{M}_2/L$ as compared to the entanglement entropy $S$ --- whereas $S$ decreases monotonically as we move from CFT to the strong coupling limit, $\mathcal{M}_2/L$ shows an initial plateau, followed by a drop after a particular coupling which we denote $g_{\star}$. This is further elucidated by considering the derivative of both quantities. 
Our results show that the crossover coupling $g_{\star}$ i.e., where magic starts to decrease sharply aligns with the coupling where the derivative of total gauge-invariant entanglement entropy has the largest magnitude. 

These two observables further align with the total particle density: $\rho_{\rm particle} = \rho_{\rm single} + \rho_{\rm double}$ where the two contributions separately count single- and double-occupancy excitations in the dressed six-state local basis [shown in Fig.~\ref{fig:six_panels} (c)]. 
Single occupancies contribute to mesonic configurations, where a quark and an anti-quark on neighboring sites are connected by a gauge link while double occupancies represent on-site baryon or anti-baryon configurations in which two quarks or two anti-quarks combine into an $\mathrm{SU}(2)$ singlet. They are a direct measure of vacuum fluctuations, and a larger value of $\rho_{\text{particle}}$ implies more local gauge-invariant fluctuations leading to more mixing of dressed-basis sectors (quantum correlations) and follows similar behavior as entanglement. The results further clarify the relation between quantum correlations (captured by entanglement and particle number density) and magic in this dressed-site $\mathrm{SU}(2)$ gauge theory. We provide additional details about the computation of all observables discussed in the main text and lower bound on non-local magic using anti-flatness~\cite{Cao:2024nrx} in Appendix~\ref{app:EE_SRE}.

\begin{figure}
\includegraphics[width=0.5\textwidth]{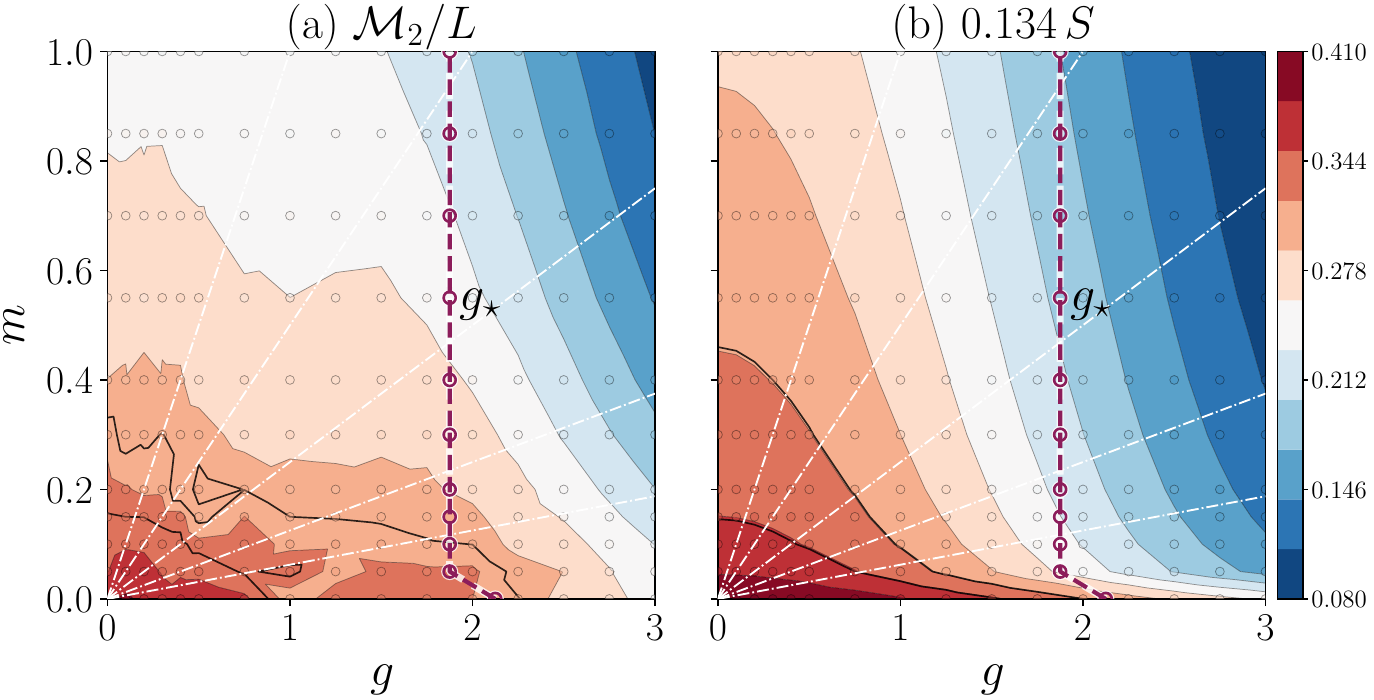}
    \caption{We show SRE normalized by lattice sites (a), total gauge-invariant entanglement entropy scaled by a constant factor to follow the same color scale (b) for $L=70$. The hollow circles denote the datasets in the $m-g$ phase diagram. The white dashed line show constant $m/g = 1, \frac{1}{2}, \frac{1}{4}, \frac{1}{8}, \frac{1}{16}$ slices. The results are obtained with a maximum bond dimension of $\chi=100$. The simulated data points are shown in hollow gray circles, and the $g_{\star}$ is shown by dashed lines.}
    \label{fig:heatmap}
\end{figure}

To further inspect the interplay between magic and entanglement entropy, we now go beyond the fixed $m=0.2$ case and compute magic and entanglement with a large lattice size, $L = 70$, for the entire $m\,-g$ plane. The results are shown in Fig.~\ref{fig:heatmap}. The white dashed lines show the different fixed dimensionless ratios of $m/g$   as one approaches the continuum limit ($g = g_{\text{dim}}a \to 0$ keeping $g_{\text{dim}}$ fixed). The islands and regions of constant color denote the region in the phase space where SRE is the same within $1\sigma$ error bars. SRE is maximum for $m=g=0$ while it decreases at different rates below and above $g_{\star} \approx 1.9$. 
For the entanglement entropy, we see a monotonic decrease as we increase $g$, but the rate of decrease is highest around $g_{\star} \approx 1.9$ (also shown in Fig.~\ref{fig:six_panels} for fixed $m=0.2$). Our results show that the location of the rapid decrease in SRE is accompanied by the steepest change with coupling in the structure of entanglement and total particle density. 

\textit{Summary and outlook} --- 
The central finding of our work --- the first large-scale study of non-stabilizerness in a non-Abelian lattice gauge theory incorporating matter fields --- is the decoupling of entanglement and magic across the confinement crossover captured by both the coupling derivative of entanglement entropy and magic. The gauge-invariant entanglement entropy decreases monotonically (but at a different rate) with increasing $g$, consistent with the standard expectation that confinement gaps the theory and suppresses long-range quantum correlations. The stabilizer R\'{e}nyi entropy, by contrast, exhibits non-monotonic behavior: it remains significant and approximately constant over an intermediate coupling regime before eventually vanishing at strong coupling. Our results point to a threshold coupling, $g_{\star}$, after which magic starts to decrease consistently. This coupling corresponds to the maximum rate of change of entanglement entropy with $g$ as shown in Fig.~\ref{fig:six_panels}. This demonstrates that non-Abelian
confinement suppresses quantum correlations (entanglement) and quantum computational complexity (magic) at different rates with the magnitude of change of entanglement entropy being maximal at the point where the non-stabilizerness starts to decrease prominently. 

We interpret the regime of sustained magic at intermediate couplings ($g \le g_{\star}$) as reflecting the structure of the confinement crossover: even as the correlation length shortens, the wavefunction retains non-trivial non-stabilizer character. The absence of a sharp peak in SRE and volume-independence of any observable at non-zero coupling $g$ is consistent with the absence of a genuine deconfinement phase transition in this $(1+1)$-dimensional model; the system is confining and in the same phase for all $g > 0$. For gauge theories that have \emph{deconfinement} transition like the $2+1$-dimensional dressed $\mathrm{SU}(2)$ gauge theory~\cite{Cataldi:2023xki}, our result suggests that both magic and derivative of entanglement entropy can be useful probes to identify singularities in the phase diagram. 

We envisage several extensions of our work. Most immediately, the effect of the irrep truncation on our results should be quantified by increasing $\jmax$ beyond $1/2$; preliminary investigation suggests the corrections are modest, but a systematic study is required which we will consider in future work. Extending the analysis to SU(3) gauge theory in $1+1$ dimensions would connect more directly to QCD-like physics. In higher dimensions, where tensor network methods face the entanglement barrier, neural quantum state ans\"{a}tze~\cite{Spriggs:2025sea} or other improved classical methods~\cite{Zhang:2025osc}
may provide a viable path forward for exploring the interplay between entanglement and magic. More broadly, the development of improved sampling algorithms for SRE and of complementary magic diagnostics 
will be essential for a complete understanding of the resource-theoretic structure of fault-tolerant quantum simulation for non-Abelian gauge theories. We leave these interesting questions for future work.

\textit{Acknowledgments} --- 
We thank Simon Catterall and Yannick Meurice for comments. R.G.J., J.I.T., B.N.B., and A.F.K. were supported by the U.S. Department of Energy, Advanced Scientific Computing Research, under contract number DE-SC0025384. G.C.T. acknowledges financial support from the National Science Foundation under award No. PHY-2325080: PIF: Software-Tailored Architecture for Quantum Co-Design.

\bibliographystyle{utphys}
%%%%%%%%%%%%%%%%%%%%%%%%%%%%%%%%%%%%%%%%
\raggedright
\bibliography{refs.bib}
%\end{document}
%%%%%%%%%%%%%%%%%%%%%%%%%%%%%%%%%%%%%%%%

\clearpage
\onecolumngrid
\appendix

\renewcommand\thefigure{S\arabic{figure}}  
\renewcommand\thetable{S\arabic{table}}  
\setcounter{figure}{0}

\onecolumngrid 
\justifying

\section{\label{app:su2_rep_gi_part}$\mathrm{SU}(2)$ representation theory and gauge singlet condition}

Let us represent group elements by matrices. 
Given a group $G$, we can associate each element $g \in G$ with a $d_j \times d_j$ matrix $D(g)$, with the property that
$D(g_1) D(g_2) = D(g_{1}g_{2})$ for any two group elements
$g_{1},g_{2} \in G$. The set of matrices $D(g)$, $g \in G$, is said to furnish a representation of $G$. The size of the matrices, $d_j$, is known as the dimension of the representation. A representation is called irreducible if it does not contain nonzero proper subrepresentations. All finite-dimensional representations of compact Lie groups are completely reducible, which means they can be written as direct sums of irreducible representations (irreps).
Moreover, all finite-dimensional representations are unitary,
i.e., $D^{\dagger}(g) D(g) = \mathbb{I}$ for all $g$ and for all representations. In particular, working with irreps of the gauge group $\mathrm{SU}(2)$ ensures that the underlying theory is unitary. For the case of the Abelian group $\mathrm{U}(1)$, all irreducible representations are one-dimensional, while for $\mathrm{SU}(2)$, the dimension is $2j+1$ for the irrep labeled by Dynkin label $j$.
As another example, in $\mathrm{SU}(3)$ gauge theory, the weight of the irreps is determined by two Dynkin labels, $p, q$, leading to the richness and exotic nature and wide range of fundamental particles. 

As shown in Fig.~\ref{fig:cartoon_dressedSU2}, each dressed site consists of left half-link, matter site, and right half-link. These have their own label $j_{L}, j_{M}, j_{R}$ corresponding to the $\mathrm{SU}(2)$ quantum numbers. The physical states are $\mathrm{SU}(2)$ singlets of the combined object at each site, i.e., $\jL \otimes \jM \otimes \jR$. When coupling three angular momenta $j_{1}$, $j_{2}$, $j_{3}$ to total $j=0$, the Clebsch-Gordan coefficient (or $3j$ symbol) is nonzero only if the following two conditions hold: (1) \emph{Triangle inequality:} $|\jL - \jM| \leq \jR \leq \jL + \jM$, and (2) \emph{Integer-sum condition:} $\jL + \jM + \jR \in \mathbb{Z}$.
The second condition is not implied by the triangle inequality alone. For example, the triple $(\jL, \jM, \jR) = (\half, \half, \half)$ satisfies the triangle inequality but has $\jL + \jM + \jR = \frac{3}{2} \notin
\mathbb{Z}$, so it does not form a gauge-singlet state. In this paper, we will restrict to $\jmax = 1/2$, which has been referred to as `hardcore gluon' in Ref.~\cite{Calajo:2024qrc}. In the main text, we specified that for a given lattice site $n$, the three $\mathrm{SU}(2)$ angular momenta entering the dressed-site construction are the left half-link flux, the matter charge, and the right half-link flux.  Up to the chosen orientation conventions, the local Gauss-law generators can be written as: 
\begin{equation}
    G_n^A
    =
    J_{L,n}^A + J_{M,n}^A + J_{R,n}^A,
    \qquad A=1,2,3 .
\end{equation}
Physical states are locally gauge invariant, and hence satisfy: 
\begin{equation}
    G_n^A \ket{\psi_{\rm phys}} = 0,
    \qquad \forall\, A=1,2,3 .
    \label{eq:Gauss_component}
\end{equation}
Equivalently, the total $\mathrm{SU}(2)$ Casimir at site $n$,
\begin{equation}
    \mathbf{G}_n^2
    \equiv
    \sum_{A=1}^{3} G_n^A G_n^A,
\end{equation}
annihilates the physical state:
\begin{equation}
    \mathbf{G}_n^2 \ket{\psi_{\rm phys}} = 0 .
    \label{eq:Gauss_Casimir}
\end{equation}
Indeed, since the generators are Hermitian, we have
\begin{equation}
    \bra{\psi_{\rm phys}} \mathbf{G}_n^2 \ket{\psi_{\rm phys}}
    =
    \sum_{A=1}^{3}
    \left\|G_n^A \ket{\psi_{\rm phys}}\right\|^2 ;
\end{equation}
hence, $\mathbf{G}_n^2=0$ is equivalent to imposing all three component
constraints $G_n^A=0$. Now $\mathbf{G}_n$ is precisely the total angular momentum obtained by
combining the three local spins,
\begin{equation}
    \mathbf{J}_{\rm tot,n}
    =
    \mathbf{J}_{L,n}
    +
    \mathbf{J}_{M,n}
    +
    \mathbf{J}_{R,n}.
\end{equation}
Therefore,
\begin{equation}
    \mathbf{G}_n^2
    =
    \mathbf{J}_{\rm tot,n}^{\,2}.
\end{equation}
A state of definite total spin $j_{\rm tot}$ obeys
\begin{equation}
    \mathbf{J}_{\rm tot,n}^{\,2}
    \ket{j_{\rm tot},m_{\rm tot}}
    =
    j_{\rm tot}(j_{\rm tot}+1)
    \ket{j_{\rm tot},m_{\rm tot}} .
\end{equation}
Hence Eq.~\eqref{eq:Gauss_Casimir} implies: 
\begin{equation}
    j_{\rm tot}(j_{\rm tot}+1)=0,
\end{equation}
which has the unique solution: 
\begin{equation}
    j_{\rm tot}=0.
\end{equation}
Thus the local Gauss-law constraint is exactly the statement that:
\begin{equation}
    j_{L,n}\otimes j_{M,n}\otimes j_{R,n}
    \mapsto 0 ,
\end{equation}
namely, the left flux, matter charge, and right flux must combine into an $\mathrm{SU}(2)$ singlet at every lattice site, i.e., 
\begin{equation}
    \bigl(j_{L,n}\otimes j_{M,n}\bigr)\otimes j_{R,n}
    \supset 0 .
\end{equation}
This is possible precisely when both the triangle law and integer condition are met. These are the local triangle/singlet conditions defining the gauge-invariant
dressed-site basis. It is straightforward to count the number of gauge-singlet states for a given truncation 
$\jmax$. Given matter occupation $n_M \in \{0, 1, 2\}$, the dimension is computed by counting all valid $(j_L, j_R)$ pairs. Empty or doubly occupied corresponds to $n_M \in \{0, 2\}$: the effective matter spin is $j_M = 0$. The triangle inequality enforces $j_L = j_R$. Since $j$ takes values from $0$ to $\jmax$ in half-integer steps, there are $(2\jmax + 1)$ valid pairs for each of the two matter states leading to  $4\jmax + 2$ states. The matter site can also be singly occupied. Then the effective matter spin is a fermion, $j_M = 1/2$. The triangle inequality enforces $j_R = j_L \pm 1/2$. Counting all valid pairs strictly within the $\jmax$ cutoff yields exactly $4\jmax$ states. Adding both, we get a total of $8\jmax + 2$ gauge-invariant states per site. This can also be computed by noting that the following combinations of irreps: 
\[
\textbf{1} \otimes \textbf{1} \otimes \textbf{1}, \;\; \textbf{2} \otimes \textbf{1} \otimes \textbf{2}, \;\; 
\textbf{2} \otimes \textbf{2} \otimes \textbf{1}, \;\;
\textbf{1} \otimes \textbf{2} \otimes \textbf{2}, \;\; \textbf{1} \otimes \textbf{1} \otimes \textbf{1},  \;\;
\textbf{2} \otimes \textbf{1} \otimes \textbf{2}
\]
each give a singlet, 
based on representation theory by
combining left half-link, matter, and right half-link 
$(\textbf{1} \oplus \textbf{2}) \, \otimes
(\textbf{1} \oplus \textbf{2} \oplus 
\textbf{1}) \, \otimes (\textbf{1} \oplus \textbf{2})$. 
These six physical states and their quantum numbers are listed in Table~\ref{tab:jmax-half-qn}. 

\begin{table}[h]
\centering
\begin{tabular}{ccccc}
\toprule
Index & $\nM$ & $j_M$ & $\jL$ & $\jR$ \\
\midrule
1 & 0 & 0        & 0        & 0        \\ \midrule
2 & 0 & 0        & $1/2$ & $1/2$ \\
\midrule
3 & 1 & $1/2$ & $0$ & $1/2$        \\
\midrule
4 & 1 & $1/2$ & $1/2$        & $0$ \\
\midrule
5 & 2 & 0        & 0        & 0        \\
\midrule
6 & 2 & 0        & $1/2$ & $1/2$ \\
\bottomrule
\end{tabular}
\caption{\label{tab:jmax-half-qn}Physical dressed-site states for irrep truncation $\jmax=1/2$. From this table, we can can deduce that $C_{L} = 2j_{L} = \text{diag}(0,1,0,1,0,1)$ and $C_{R} = 2j_{R} = \text{diag}(0,1,1,0,0,1)$.}
\end{table}

\section{\label{app:sec_H_cons}Dressed basis and the construction of Hamiltonian}

\emph{Dressed basis} --- As mentioned in the previous section, each dressed site in the effective irrep-truncated basis is labeled by three quantum numbers. These are left half-link carrying irrep $\jL \in \{0, \half, 1, \ldots, \jmax\}$, matter site with occupation $\nM \in \{0,1,2\}$ and matter irrep
$\jM = 0$ for $\nM=0,2$ and $\jM=\half$ for $\nM=1$, and
right half-link carrying irrep $\jR \in \{0, \half, 1, \ldots, \jmax\}$. The local basis is therefore a \emph{dressed-site half-link basis} $\ket{\nM,\jL,\jR}$, and gauge invariance is imposed locally by requiring the three irreps $\jL \otimes \jM \otimes \jR$ to couple to a singlet. We have a total of 36 (physical + unphysical) states for each dressed site given by:
\begin{equation}
 (\textbf{1} \oplus \textbf{2}) \, \otimes
 (\textbf{1} \oplus \textbf{2} \oplus 
 \textbf{1}) \, \otimes (\textbf{1} \oplus \textbf{2}).
\end{equation}
However, only gauge singlets states are physical. There are six of these. For hardcore gluon truncation, i.e., $\jmax=1/2$, each lattice site has six physical (gauge-invariant) states with the quantum numbers as written down in Table.~\ref{tab:jmax-half-qn}. Using the notation that $\ket{\cdot,\cdot,\cdot}$ denotes
\text{(left half-link sector, matter sector, right half-link sector)}, we have: 
\begin{align}
\ket{1} &= |0,0,0\rangle, \label{eq:state1} \\[4pt]
|2\rangle &= \frac{|\textcolor{red}{r},0,\textcolor{green}{g}\rangle - |\textcolor{green}{g},0,\textcolor{red}{r}\rangle}{\sqrt{2}}, \label{eq:state2} \\[4pt]
|3\rangle &= \frac{|\textcolor{green}{g},\textcolor{red}{r},0\rangle - |\textcolor{red}{r},\textcolor{green}{g},0\rangle}{\sqrt{2}},\label{eq:state3} \\[4pt]
|4\rangle &= \frac{|0,\textcolor{red}{r},\textcolor{green}{g}\rangle - |0,\textcolor{green}{g},\textcolor{red}{r}\rangle}{\sqrt{2}}, \label{eq:state4} \\[4pt]
|5\rangle &= |0,d,0\rangle, \label{eq:state5} \\[4pt]
|6\rangle &= \frac{|\textcolor{red}{r},d,\textcolor{green}{g}\rangle - |\textcolor{green}{g},d,\textcolor{red}{r}\rangle}{\sqrt{2}} \label{eq:state6},
\end{align}
which correspond to $\textbf{1} \otimes \textbf{1} \otimes \textbf{1}$, $\textbf{2} \otimes \textbf{1} \otimes \textbf{2}$, $\textbf{2} \otimes \textbf{2} \otimes \textbf{1}$, 
$\textbf{1} \otimes \textbf{2} \otimes \textbf{2}$, $\textbf{1} \otimes \textbf{1} \otimes \textbf{1}$ and 
$\textbf{2} \otimes \textbf{1} \otimes \textbf{2}$, respectively.

Note that if we had $\jmax = 1$, i.e., 
$(\textbf{1} \oplus \textbf{2} \oplus  \textbf{3}) \, \otimes
 (\textbf{1} \oplus \textbf{2} \oplus 
 \textbf{1}) \, \otimes (\textbf{1} \oplus \textbf{2} \oplus  \textbf{3})$, 
we will have 10 singlets in total. In addition to the six singlets from $\jmax=1/2$, we will have singlet contributions from $\textbf{2} \otimes \textbf{2} \otimes \textbf{3}$,  $\textbf{3} \otimes \textbf{2} \otimes \textbf{2}$,  $\textbf{3} \otimes \textbf{1} \otimes \textbf{3}$, and $\textbf{3} \otimes \textbf{1} \otimes \textbf{3}$. 

\emph{Hamiltonian construction} --- We have three terms in the Hamiltonian. The mass and electric field terms can be written by defining four diagonal operators from the quantum numbers of each physical state:
\begin{align}
    M  &= \operatorname{diag}(\nM), \label{eq:M_def} &
    C  &= C_L + C_R = \operatorname{diag}(2\jL + 2\jR), \\
    D_L &= \operatorname{diag}\bigl((-1)^{2\jL}\bigr) = \text{diag}(+1, -1, +1, -1, +1, -1), &
    D_R &= \operatorname{diag}\bigl((-1)^{2\jR}\bigr) = \text{diag}(+1, -1, -1, +1, +1, -1) \label{eq:D_def} .
\end{align}
Here $M$ is the matter number operator, $C$ (or $C_2$) is the quadratic Casimir of a single link (proportional to $j(j+1)$, the electric energy of the link), and $D_{L,R}$ are parity operators that are useful diagnostics for monitoring the bond matching condition. These are all represented by finite $\dphys\times\dphys$ matrices. 
Although the on-site Gauss's law is built into the dressed-site basis, the inter-site link constraint $\jR^{(n)} = \jL^{(n+1)}$ is not automatically enforced. We monitor this during our simulations by computing: 
\begin{equation}
    V_b = \sum_{j \neq j'}
    \langle\psi|\, P^{(b-1)}_{\jR = j}\, P^{(b)}_{\jL = j'}\, |\psi\rangle,
\end{equation}
where $P^{(n)}_{\jR = j}$ projects site $n$ onto states with right-half-link irrep $j$. A physical state has $V_b = 0$ on every bond and any non-zero value
indicates leakage out of the gauge-invariant subspace. In all our simulations, we have checked that $V_b = 0$, which ensures that all the physics we extract is gauge-invariant. In case this is not satisfied, likely for higher $\jmax$ truncation, one can add a penalty term to ensure this. 
The penalty term can be written as:   
\begin{equation}
    H_{\rm pen}
    = \lambda \sum_n \bigl(C_R^{(n)} - C_L^{(n+1)}\bigr)^2.
\end{equation}
Since $C_R=2\jR$ and $C_L=2\jL$, this penalizes any violation of the half-link matching condition across a bond. Numerically, it helps keep simulations inside the gauge-invariant sector even when truncation errors are present.

For the electric field term, note that we have a slightly modified definition compared to the standard KS Hamiltonian similar to Ref.~\cite{Calajo:2024qrc}. In the canonical Hamiltonian, the electric energy is a sum over \emph{physical links}
written as: 
\begin{equation}
H_{\rm elec}^{\rm can}
=
\frac{g_{\rm can}^2}{2}\sum_{\ell} E_\ell^2
=
\frac{g_{\rm can}^2}{2}\sum_{\ell} j_\ell(j_\ell+1),
\end{equation}
where $j_\ell$ is the $\mathrm{SU}(2)$ irrep carried by link $\ell$. In the present work, we restrict to the hardcore truncation $j_{\max}=1/2$, so each link can only carry $j_\ell=0$ or $j_\ell=1/2$. On this restricted Hilbert space, one can replace the quadratic Casimir by the linear projector onto the nontrivial flux sector: 
\begin{equation}
\sum_{j_l} j_\ell(j_\ell+1) = \sum_{j_l} \frac{3}{4}\,(2j_\ell),
\qquad j_\ell\in\left\{0,\frac12\right\}.
\label{eq:casimir_linear_relation}
\end{equation}
In the dressed-site basis, we use $g^2 \sum_n C_n$, and the canonical electric term may be written as $H_{\rm elec}=\frac{g_{\rm can}^2}{2}\sum_{\ell} j_\ell(j_\ell+1)
=
\frac{3g_{\rm can}^2}{8}\sum_n C_n$. 
In our notation, the electric term is simply written as $g^2 \sum_{n} C_{n}$ [Eq.~\eqref{eq:H_C_n_appendix}].

We find that to construct the Hamiltonian, we need
hopping matrices in addition to the diagonal matrices defined in Eqs.~\eqref{eq:M_def}. The hopping operators are~\cite{Calajo:2024qrc}: 
\begin{align}
\label{eq:A1}
\hAone &=
\begin{pmatrix}
0 & 0 & 0 & \sqrt2 & 0 & 0 \\
0 & 0 & 1 & 0 & 0 & 0 \\
0 & 1 & 0 & 0 & 0 & 1 \\
\sqrt2 & 0 & 0 & 0 & \sqrt2 & 0 \\
0 & 0 & 0 & \sqrt2 & 0 & 0 \\
0 & 0 & 1 & 0 & 0 & 0
\end{pmatrix},
&
\hAtwo &=
i\begin{pmatrix}
0 & 0 & 0 & \sqrt2 & 0 & 0 \\
0 & 0 & 1 & 0 & 0 & 0 \\
0 & -1 & 0 & 0 & 0 & 1 \\
-\sqrt2 & 0 & 0 & 0 & \sqrt2 & 0 \\
0 & 0 & 0 & -\sqrt2 & 0 & 0 \\
0 & 0 & -1 & 0 & 0 & 0
\end{pmatrix},
\\[8pt]
\label{eq:B1}
\hBone &=
i\begin{pmatrix}
0 & 0 & -\sqrt2 & 0 & 0 & 0 \\
0 & 0 & 0 & -1 & 0 & 0 \\
\sqrt2 & 0 & 0 & 0 & -\sqrt2 & 0 \\
0 & 1 & 0 & 0 & 0 & -1 \\
0 & 0 & \sqrt2 & 0 & 0 & 0 \\
0 & 0 & 0 & 1 & 0 & 0
\end{pmatrix},
&
\hBtwo &=
\begin{pmatrix}
0 & 0 & \sqrt2 & 0 & 0 & 0 \\
0 & 0 & 0 & 1 & 0 & 0 \\
\sqrt2 & 0 & 0 & 0 & \sqrt2 & 0 \\
0 & 1 & 0 & 0 & 0 & 1 \\
0 & 0 & \sqrt2 & 0 & 0 & 0 \\
0 & 0 & 0 & 1 & 0 & 0
\end{pmatrix}.
\end{align}
All four are Hermitian. They descend from two non-Hermitian operators
$\hat{Q}_L$, $\hat{Q}_R$ through: 
\begin{equation}
\label{eq:AB_from_Q}
\hat{A}^{(1)} = \hat{Q}_L + \hat{Q}_L^\dagger,\quad
\hat{A}^{(2)} = i(\hat{Q}_L - \hat{Q}_L^\dagger),\qquad
\hat{B}^{(1)} = \hat{Q}_R + \hat{Q}_R^\dagger,\quad
\hat{B}^{(2)} = i(\hat{Q}_R - \hat{Q}_R^\dagger),
\end{equation}
with inverses
\begin{equation}
\label{eq:Q_from_AB}
\hat{Q}_L = \tfrac{1}{2}(\hat{A}^{(1)} - i\hat{A}^{(2)}),\qquad
\hat{Q}_R = \tfrac{1}{2}(\hat{B}^{(1)} - i\hat{B}^{(2)}).
\end{equation}
Physically,
$\hat{Q}_L$ annihilates matter and acts on $j_L$, while
$\hat{Q}_R$ annihilates matter and acts on $j_R$. The hopping can be rewritten as
\begin{equation}
\label{eq:hopQ}
\hat{A}^{(1)}_n\hat{B}^{(1)}_{n+1}
+\hat{A}^{(2)}_n\hat{B}^{(2)}_{n+1}
= 2\bigl(\hat{Q}_{L,n}^\dagger\,\hat{Q}_{R,n+1}
        +\hat{Q}_{L,n}\,\hat{Q}_{R,n+1}^\dagger\bigr).
\end{equation}

Concretely, these hopping operators are better understood as being obtained from $\mathrm{SU}(2)$ angular-momentum recoupling and can be expressed in terms of Wigner-$6j$ symbol which is more amenable to general arbitrary $\jmax$ truncation. In other words, a more general expression of the $Q$ operators for arbitrary finite truncation $\jmax$ can be written in terms of Wigner-$6j$ symbols. Due to the constraint imposed by the Gauss's law, the system can be viewed as the coupling of two angular momenta, where the matter field effectively arises as the sum of  these two angular momentum contributions. For a composite system of two angular momenta $j_1$ and $j_2$ coupled to total $j$, if a tensor operator $T^{(k)}$ acts only on the $j_1$ subsystem, then the corresponding reduced matrix element can be written as \cite{edmonds1996angular}: 
\begin{equation}   
(y j_1 j_2 j ||T^{(\kappa)} || y' j_1' j_2 j') = (-1)^{j_1 + j_2 + j' + \kappa} \sqrt{(2j+1)(2j'+1)} \begin{Bmatrix} j_1 & j & j_2 \\ j' & j_1' & \kappa \end{Bmatrix} (y j_1 || T^{(\kappa)} || y' j_1').
\end{equation}
Applying this identity to our physical state basis allows us to derive the explicit matrix elements for the link-variable $Q$ operators
\begin{equation}
    \langle n_M+1,\, j_L', \, j_R |\hat{Q}^\dagger_L|n_M, j_L, j_R\rangle = \sqrt{2} \, (-1)^{2(j_R+j_L+j_M')} \sqrt{(2j_M+1)(2j_M'+1)} \, \sixj{j_L'}{j_M'}{j_R}{j_M}{j_L}{\frac{1}{2}},  
\end{equation}
and 
\begin{equation}
    \langle n_M+1,\, j_L, \, j_R' |\hat{Q}^\dagger_R|n_M, j_L, j_R\rangle = i\sqrt{2} \, (-1)^{2(j_R+j_L+j_M')} \sqrt{(2j_M+1)(2j_M'+1)} \, \sixj{j_R'}{j_M'}{j_L}{j_M}{j_R}{\frac{1}{2}}.  
\end{equation}
By taking linear combinations of these two expressions as prescribed by Eq.\ \eqref{eq:Q_from_AB}, we obtain the matrix elements for the $\hat{A}$ and $\hat{B}$ operators: 
\begin{equation}
    \langle n_M+1,\, j_L', \, j_R|\hat{A}^{(1)}|n_M,\, j_L, \, j_R\rangle=\sqrt{2}\,(-1)^{2(j_R+j_L+j_M')}\,  \sqrt{(2j_M+1)(2j'_M+1)}\, \begin{Bmatrix}
j_L' & j_M' & j_R\\
j_M & j_L &  1/2
\end{Bmatrix}, 
\end{equation}
and
\begin{equation}
    \langle n_M,\, j_L', \, j_R|\hat{A}^{(1)}|n_M+1,\, j_L, \, j_R\rangle=\sqrt{2}\,\big((-1)^{2(j_R+j_L'+j_M)}\big)^*\,  \sqrt{(2j_M+1)(2j'_M+1)}\, \begin{Bmatrix}
j_L' & j_M' & j_R\\
j_M & j_L & 1/2 
\end{Bmatrix}. 
\end{equation}
Following the same logic for the second  component $\hat{A}^{(2)}$, we find:
\begin{equation}
    \langle n_M+1,\, j_L', \, j_R|\hat{A}^{(2)}|n_M,\, j_L, \, j_R\rangle=-i\sqrt{2}\,(-1)^{2(j_R+j_L+j_M')}\,  \sqrt{(2j_M+1)(2j'_M+1)}\, \begin{Bmatrix}
j_L' & j_M' & j_R\\
j_M & j_L &  1/2 
\end{Bmatrix},
\end{equation}
and
\begin{equation}
    \langle n_M,\, j_L', \, j_R|\hat{A}^{(2)}|n_M+1,\, j_L, \, j_R\rangle=i\sqrt{2}\,\big((-1)^{2(j_R+j_L'+j_M)}\big)^*\,  \sqrt{(2j_M+1)(2j'_M+1)}\, \begin{Bmatrix}
j_L' & j_M' & j_R\\
j_M & j_L & 1/2 
\end{Bmatrix}.
\end{equation}
Similarly, substituting the explicit forms of $Q_R$ into Eq.~\eqref{eq:Q_from_AB} yields the matrix expressions for the $B$ operators. For $B^{(1)}$, the elements are:
\begin{equation}
    \langle n_M+1,\, j_L, \, j_R' |\hat{B}^{(1)}|n_M, j_L, j_R\rangle = i\sqrt{2} \, (-1)^{2(j_R+j_L+j_M')} \sqrt{(2j_M+1)(2j_M'+1)} \, \sixj{j_R'}{j_M'}{j_L}{j_M}{j_R}{\frac{1}{2}}, 
\end{equation}
and 
\begin{equation}
    \langle n_M,\, j_L, \, j_R'|\hat{B}^{(1)}|n_M+1,\, j_L, \, j_R\rangle=-i\sqrt{2} \, \big((-1)^{2(j_R'+j_L+j_M)}\big)^* \sqrt{(2j_M+1)(2j_M'+1)} \, \sixj{j_R'}{j_M'}{j_L}{j_M}{j_R}{\frac{1}{2}}.  
\end{equation}
Finally,  for $\hat{B}^{(2)}$:  
\begin{equation}
    \langle n_M+1,\, j_L, \, j_R '|\hat{B}^{(2)}|n_M, j_L, j_R\rangle = \sqrt{2} \, (-1)^{2(j_R+j_L+j_M')} \sqrt{(2j_M+1)(2j_M'+1)} \, \sixj{j_R'}{j_M'}{j_L}{j_M}{j_R}{\frac{1}{2}}, 
\end{equation}
and
\begin{equation}
    \langle n_M,\, j_L, \, j_R'|\hat{B}^{(2)}|n_M+1,\, j_L, \, j_R\rangle=\sqrt{2} \, \big((-1)^{2(j_R'+j_L+j_M)}\big)^* \sqrt{(2j_M+1)(2j_M'+1)} \, \sixj{j_R'}{j_M'}{j_L}{j_M}{j_R}{\frac{1}{2}}. 
\end{equation}
Using all these, the full Hamiltonian (without the penalty term) is given as:
\begin{equation}
H = H_{\mathrm{mass}} + H_{\mathrm{elec}} + H_{\mathrm{hop}},
\label{eq:Ham_2}
\end{equation}
with
\begin{align}
H_{\mathrm{mass}} &= m\sum_n (-1)^n M_n,
\\
H_{\mathrm{elec}} &= g^2 \sum_n C_n, \label{eq:H_C_n_appendix}
\\
H_{\mathrm{hop}} &= \sum_n
\Bigl[\hAone_{n} \hBone_{n+1} + \hAtwo_{n} \hBtwo_{n+1}
\Bigr],
\end{align}
where $M_n$, $C_n$, $\hAone$, $\hAtwo$, $\hBone$, and 
$\hBtwo$ are defined in Eqs.~\eqref{eq:M_def}, \eqref{eq:A1}, and \eqref{eq:B1}, respectively.

\section{\label{app:EE_SRE}Additional details about the observables}

\subsection{Entanglement entropy}

\begin{figure}
    \centering
    \includegraphics[width=1.0\linewidth]{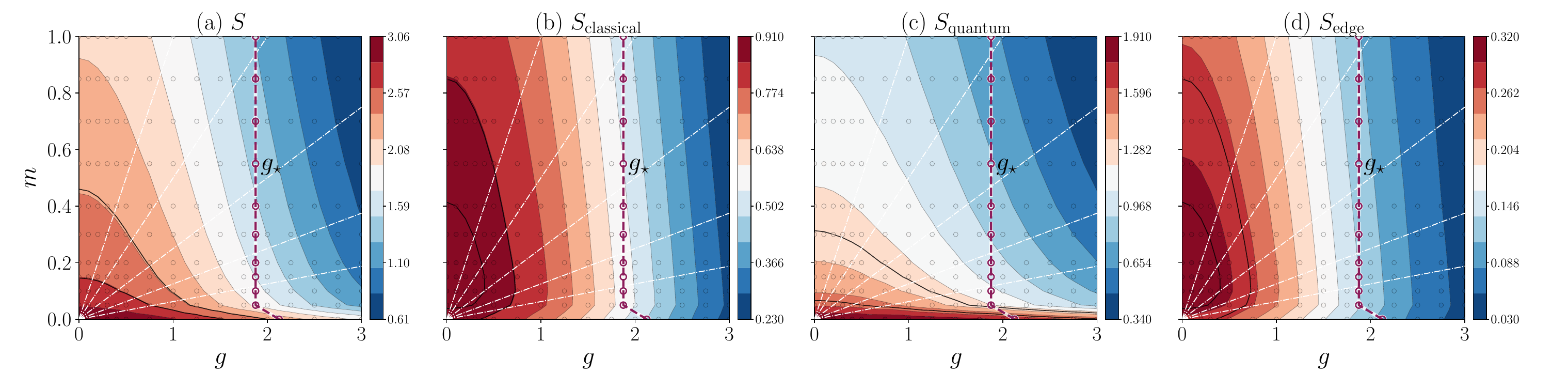}
    \caption{The different components of the total entropy for $L=70$. The simulated data points are shown in hollow gray circles. The dashed line shows $g_{\star}$, which is computed where the rate of change of $S$ with $g$ has largest magnitude.}
    \label{fig:app_four_panels}
\end{figure}

In lattice gauge theories, computing the entanglement entropy, $S_{\text{EE}}$ across a spatial bipartition (into regions $A$ and $\bar{A}$) is subtle. The standard recipe for the bipartite entanglement entropy of a region $A$ in a quantum spin chain is: 
\begin{equation}
  S_A = -\Tr\bigl[\rhoA \ln \rhoA\bigr],
  \qquad \rhoA = \Tr_{\bar A} \ket{\psi}\bra{\psi},
\end{equation}
which assumes a tensor-product factorization $\mathcal{H} = \HA \otimes \mathcal{H}_{\bar A}$. 
In lattice gauge theories this factorization fails: the irrep label $\jcut$ on a link cut by the
boundary $\partial A$ appears in the Gauss law on both sides of the
cut, so basis states cannot be split freely into $A$ and  
$\bar A$. In $\mathrm{SU}(2)$ lattice gauge theory, the boundary flux is given by the irrep $j_{\text{cut}}$. 
If we cut the bond between sites $n-1$ and $n$, the relevant boundary irrep is 
the outgoing right link of the left site: $j_{\text{cut}} = j_R^{(n-1)}$ (shown in Fig.~\ref{fig:EE_cut_fig}).
Because the global state satisfies the link law: $j_R^{(n-1)} = j_L^{(n)}$, 
the reduced density matrix $\rho_A$ is block-diagonal with respect to the boundary irrep $j_{\text{cut}}$:
\begin{equation}
    \rho_A = \bigoplus_{j_{\text{cut}}} p_{j_{\text{cut}}} ~\rho_{A, j_{\text{cut}}},
\end{equation}
where $p_{j_{\text{cut}}}$ is the probability of the boundary link taking the value $j_{\text{cut}}$, and $\rho_{A, j_{\text{cut}}}$ is the normalized 
reduced density matrix restricted to that specific flux sector. In our case, since we use $\jmax = 1/2$, the possible values for $j_{\text{cut}}$ are $0, 1/2$. To compute the true gauge-invariant entanglement entropy, one must partition the reduced density matrix $\rho_A$ into superselection sectors labeled by the boundary flux $j_{\text{cut}}$.
The issue of edge modes and entanglement entropy has been the subject of intense investigations~\cite{Balachandran:1994vi, Buividovich:2008gq, Donnelly:2008vx, Donnelly:2011hn, Casini:2013rba}. We follow the prescription of Ref.~\cite{Donnelly:2011hn} on entanglement in gauge theories~\footnote{One of us (R.G.J.) would like to thank Will Donnelly for discussions on this during Fall 2019} where the total gauge-invariant von Neumann entanglement entropy decomposes into three distinct, physically meaningful parts (referring to $j_{\text{cut}}$ simply as $j$): 
\begin{equation}
    S = \underbrace{-\sum_j p_j \ln p_j}_{S_{\text{classical}}} + \underbrace{\sum_j p_j S_j}_{S_{\text{quantum}}} + \underbrace{\sum_j p_j \ln(d_j)}_{S_{\text{edge}}}. 
\end{equation}
The first term measures the classical fluctuation (Shannon entropy) of the electric flux across the entanglement cut. The second term is the weighted average of the internal quantum entanglement within each superselection sector where $S_{\text{quantum}} = -\Tr(\rho_{A,j} \ln \rho_{A,j})$.
The third term takes care of the boundary/edge term~\cite{Donnelly:2008vx, Casini:2013rba} corresponding to the internal degeneracy of the gauge flux on the boundary.
For an Abelian gauge theory such as quantum electrodynamics $\mathrm{U}(1)$, all irreducible representations are one-dimensional ($d_j = 1$) leading to $S_{\text{edge}}^{\mathrm{U}(1)} = 0$. 
For non-Abelian gauge theory like $\mathrm{SU}(2)$ and $\mathrm{SU}(3)$, this term depends on the dimension of the irreducible representation (irrep) of the gauge group labeled by Dynkin labels. We show the composition of the full entanglement entropy for $L=70$ in the $m-g$ plane in Fig.~\ref{fig:app_four_panels}.

\subsection{\label{subsec:SRE}Stabilizer R\'{e}nyi entropy as a probe of magic}

In general, quantifying magic is not a straightforward task, but in recent years the stabilizer R\'{e}nyi entropy (SRE)~\cite{Leone:2021rzd} has emerged as one of the most widely used magic measures, with several approaches available for its computation.
One such approach is the replica method~\cite{Haug:2022vpg}; however, it scales exponentially with system size, rendering the computation of magic for large systems intractable. Instead, we follow an alternative approach based on importance-sampling Monte Carlo methods~\cite{Tarabunga:2023ggd, Tarabunga2023ManyBodyMagic, Lami:2023naw}. 
The sampling approach to SRE proceeds by introducing 
a probability distribution over the exponentially large set of Pauli strings:
\begin{equation}
q(P) \;=\; \frac{\bigl|\langle P\rangle\bigr|^{2}}{Z},
\qquad
Z \;=\; \sum_{P\in\mathcal{P}_N}\bigl|\langle P\rangle\bigr|^{2}.
\label{eq:q_dist}
\end{equation}
For a pure state, the Pauli 2-design identity gives
$Z = 2^{N}$,
so that $q(P)$ is properly normalized without explicit computation of the
partition function.
The order-2 SRE can then be written as an expectation value under this
distribution:
\begin{align}
\mathcal{M}_{2}
&= -\log_{2}\!\Bigg(\frac{1}{2^{N}}
\sum_{P\in\mathcal{P}_N}
\bigl \vert \langle P\rangle\bigr \vert  ^{4}\Bigg)
= -\log_{2} \Big(\mathbb{E}_{P \sim q}\,
\bigl \vert  \langle P\rangle\bigr \vert^{2}\Big),
\label{eq:sre2}
\end{align}
which is estimated by Monte Carlo sampling from $q(P)$ and averaging the squared expectation values $|\langle P\rangle|^{2}$ of the drawn strings.
Each dressed site has physical dimension $d=6$ (for $j_{\max}=\tfrac{1}{2}$)
and is embedded into $n_q = \lceil\log_2 d\rceil = 3$ qubits via the trivial
isometry $V|i\rangle = |i\rangle$ for $i=0,\ldots,d{-}1$.
At each site we construct all $4^{n_q} = 64$ three-qubit Pauli strings and project
each onto the physical subspace by taking its upper-left $d\times d$ block.
These 64 projected matrices are precomputed and stored as a contiguous array,
then reused across all scan points. The total number of qubits required to simulate an $L$-site Hamiltonian is $N = 3L$.
Since a global Pauli string factorizes across MPS sites, $P = P^{(1)}\otimes\cdots\otimes P^{(L)}$,
its expectation value $\langle P\rangle$ is computed by a left-to-right
transfer-matrix contraction at cost $O(L\chi^{2}d)$ per string, where $\chi$
is the MPS bond dimension.
To sample from $q(P)$, we use a single-site Gibbs sampler that exploits this
factorized structure. A global Pauli string assigns one of 64 local Pauli string $P_{k_n}$ to each
site~$n$, so that we can keep their information as
$\mathbf{k} = (k_1,\ldots,k_L)$.
In each Gibbs update at site~$n$, we hold all other assignments
$k_{m\neq n}$ fixed and resample $k_n$ from the conditional distribution
\begin{equation}
q(k_n \mid \mathbf{k}_{\setminus n})
\;\propto\;
\bigl|\langle P_{\mathbf{k}}\rangle\bigr|^{2}.
\label{eq:gibbs_conditional}
\end{equation}
Because the MPS contraction factorizes across sites, the dependence on $k_n$
enters only through the local tensor at site~$n$.
All contributions from sites to the left and right can be absorbed into
environment tensors $E_{n-1}$ and $R_{n+1}$, which encode the partial
contraction of the bra--ket network up to site~$n{-}1$ and from
site~$n{+}1$ onward, respectively.
Contracting these environments with the MPS tensor at site~$n$ (but leaving
the physical indices open) yields a $d\times d$ matrix $M^{(n)}$, from which
the conditional probability for each local Pauli $P_k$ is
\begin{equation}
q(k \mid \mathbf{k}_{\setminus n})
\;\propto\;
\bigl|\operatorname{Tr}[M^{(n)}\, P_k]\bigr|^{2}.
\label{eq:cond_from_M}
\end{equation}
All 64 values are obtained in one tensor contraction of $M^{(n)}$ against
the cached stack of projected Pauli matrices, and a new $k_n$ is drawn
from the resulting categorical distribution. The key algorithmic optimization is environment caching: rather than
recomputing the full $O(L\chi^{2}d)$ contraction for every candidate Pauli leading to 
total complexity of $O(L^2\chi^{2}d \,4^{\log_{2}(d)}) $,
the left environments $\{E_n\}$ are built incrementally during a forward
sweep (sites $1$ to $L$) and the right environments $\{R_n\}$ during a
backward sweep ($L$ to $1$). In addition, the tensor contraction paths for the environment updates are computed once and cached, eliminating repeated path-optimization overhead. This leads to improvement of complexity to 
$O(L(\chi^{3}d + 4^{\log_{2}(d)}d^{2}))$, which becomes important for scaling to large $L$ as done in this work. For our case with $d=6$ and $\chi=100$, the leading complexity is simply
$\widetilde{O}(L\chi^{3}d)$. We computed SRE using tensor networks using open-source software packages i\textsc{Tensor}~\cite{Fishman_2022} and \textsc{TenPy}~\cite{Hauschild_2024} and compared the results 
between the two implementations.

We typically collected 3000 samples after a thermalization cut of about $300$ samples. The dependence of the estimator on the number of samples was found to be
negligible within error bars. For $L=50$ with $m=0.2$ and $g=0.05$, we obtained $\mathcal{M}_{2} = 15.75(22)$ with 2000 samples and $\mathcal{M}_{2} = 15.82(13)$ with 3000 samples where the statistical errors were estimated using bootstrap methods. As a consistency check, we verify that $\mathcal{M}_{2}/L$ is independent of
$L$ at fixed parameters with $m, g \neq 0$, confirming that the computed SRE
is extensive. A further check is that $\mathcal{M}_{2}\to 0$ in the strong-coupling limit
$g\to\infty$, where the ground state approaches a product state.
In our results, we mostly report $\mathcal{M}_{2}/L$ alongside the entanglement
entropy density $S/L$, since both are intensive quantities. Away from the CFT limit, we can fit the SRE to a two-parameter fit and extract exponents $\alpha$ and $\gamma$ at fixed $m=0.2$. The results for $m=0.2$ with $L = 20, 30, 40, 50$ are presented in Table~\ref{tab:alpha_gamma_m02} below.

\begin{table}
\centering
\label{tab:alpha_gamma_fit_short}
\begin{tabular}{ccc}
\hline\hline
$g$ & $\alpha$ & $\gamma$ \\
\hline
1.0 & 0.208(3) & -0.289(73) \\
2.0 & 0.198(2) & -0.113(55) \\
3.0 & 0.1415(8) & -0.081(26) \\
5.0 & 0.0583(3) & -0.0071(91) \\
10.0 & 0.00961(4) & -0.0042(14) \\
\hline\hline
\end{tabular}
\caption{\label{tab:alpha_gamma_m02}Two-parameter fits of the stabilizer R\'enyi entropy to $\mathcal{M}_2(L)=\alpha L+\gamma$ for $L=20,30,40,50$ at fixed $m=0.2$.
Here the logarithmic coefficient of Eq.~\eqref{eq:M2-CFT-pred} is set to $\beta=0$. The decrease of both $\alpha$ and $\gamma$ toward zero at large $g$ is consistent with the approach to a zero magic strong-coupling state.}
\end{table}

\subsection{Total particle density}\label{app:Baryon_density}
We study vacuum fluctuations through the total particle density:
$\rho_{\rm particle} = \rho_{\rm single} + \rho_{\rm double}$,
where the two contributions separately count single- and double-occupancy excitations in the dressed six-state local basis. 
The single-occupancy contribution is obtained from the local probabilities of the dressed states
$\ket{3}$ and $\ket{4}$ (see Eqs.~\eqref{eq:state3}, ~\eqref{eq:state4}),
\begin{equation}
    p_{\rm single}(n)
    =
    \big\langle P^{(3)}_n \big\rangle
    +
    \big\langle P^{(4)}_n \big\rangle,
    \label{eq:b_single}
\end{equation}
where $P^{(a)}_n=\ket{a}\!\bra{a}$ is the projector onto the $a^{\text{th}}$ local dressed state at site $n$. The double-occupancy depends on the choice of reference Dirac sea. Using the lattice staggered vacuum as:   
\begin{equation}
    \ket{\Omega}
    =
\underbrace{\ket{1}}_{\text{odd}}\ket{5}\ket{1}\ket{5}\cdots,
\end{equation}
the double baryon density is defined as: 
\begin{equation}
    p_{\rm double}(n)
    =
    \begin{cases}
    \big\langle P^{(5)}_n \big\rangle
    +
    \big\langle P^{(6)}_n \big\rangle,
    & n\ \text{odd},
    \\[2mm]
    \big\langle P^{(1)}_n \big\rangle
    +
    \langle P^{(2)}_n \rangle
    ,
    & n\ \text{even}.
    \end{cases}
    \label{eq:b_double}
\end{equation}
Then, using Eqs.~\eqref{eq:b_single} and \eqref{eq:b_double}, we can define: 
\begin{equation}
    \rho_{\rm particle}
    =
    \frac{1}{L}
    \sum_{n=1}^{L}
    \left[
    p_{\rm single}(n)
    +
    p_{\rm double}(n)
    \right],
\end{equation}
as shown in Fig.~\ref{fig:six_panels} for different $L$ with fixed $m=0.2$.

\subsection{Anti-flatness}

\begin{figure}
    \centering
    \includegraphics[width=0.7\linewidth]{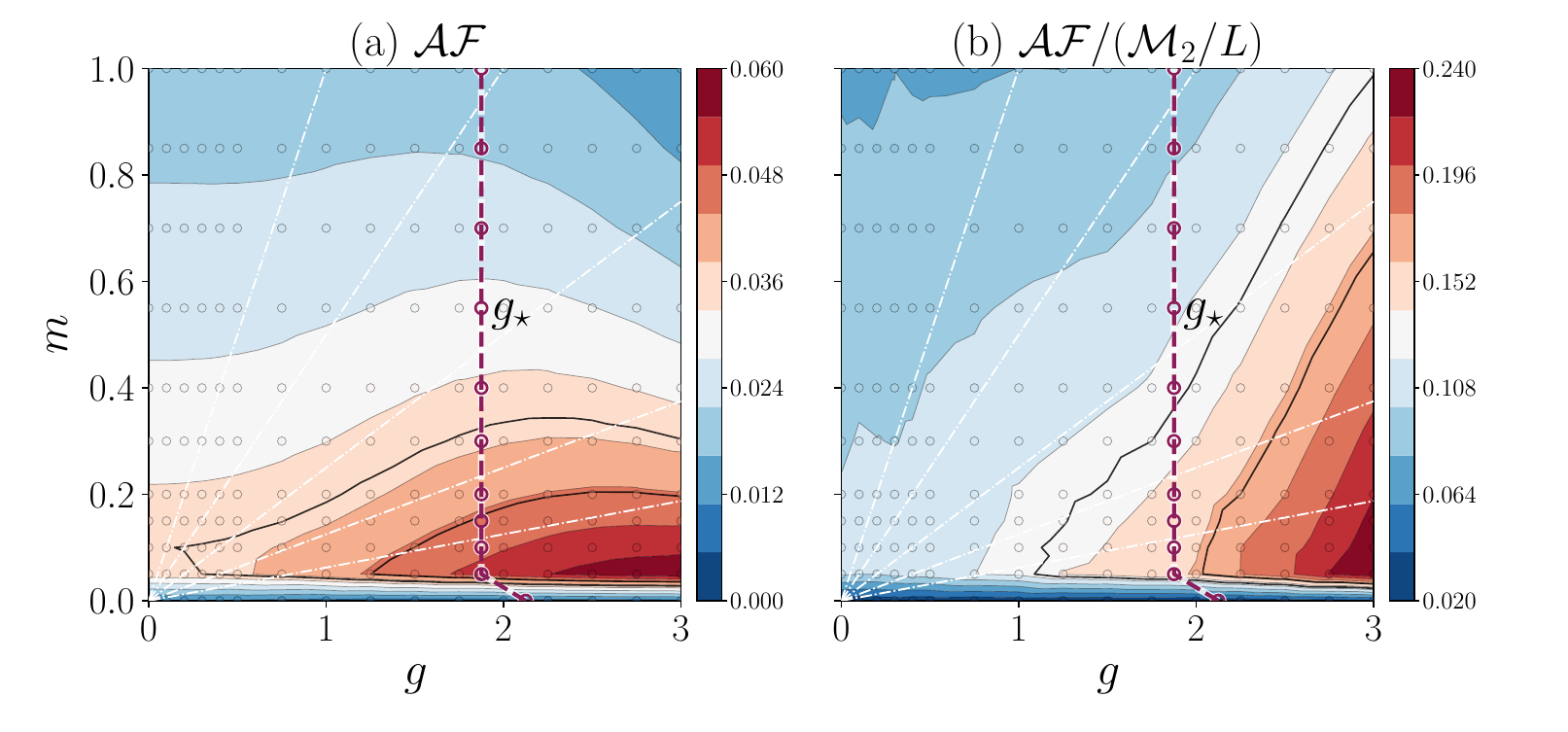}
    \caption{We show the behavior of anti-flatness and the ratio of anti-flatness with SRE for $L = 70$.}
    \label{fig:AF_and_its_ratio_with_SRE}
\end{figure}

An additional observable that we compute for the $\mathrm{SU}(2)$ gauge theory with matter is the deviation from flatness of the entanglement spectrum.
This quantity has been useful for characterizing the complexity of quantum states and for putting bounds on the non-local magic. 
Unlike the full SRE that we computed in the main text, it is sensitive to the correlations across the boundary. We define the anti-flatness of a reduced density matrix $\rho_A$ as: 
\begin{equation}\label{eq:AF}
\mathcal{AF} = \Tr(\rho_A^3) - \bigl[\Tr(\rho_A^2)\bigr]^2 \,.
\end{equation}
This quantity was introduced as a subsystem-level diagnostic of non-Gaussianity and non-stabilizerness of quantum states~\cite{Jasser:2026hrg}.
The name arises from the observation that $\mathcal{AF}(\rho_A) = 0$ if and only if the entanglement spectrum $\{\lambda_i\}$ of $\rho_A$ is \emph{flat} within its support, 
i.e., $\rho_A$ is proportional to a projector. Expressing $\Tr(\rho_A^n) = \sum_i \lambda_i^n$ in terms of the eigenvalues $\lambda_i$ of $\rho_A$ (with $\sum_i \lambda_i = 1$), the anti-flatness can be re-expressed as
\begin{equation}
\mathcal{AF} = \sum_i \lambda_i^3 - \Bigl(\sum_i \lambda_i^2\Bigr)^2 \,.
\end{equation}
Note that we have $\mathcal{AF}=0$ for product states and for states with a flat reduced spectrum. 
In general, anti-flatness is generically non-zero 
for non-stabilizer states with non-trivial entanglement. The connection to magic is that stabilizer states, when bi-partitioned, always produce reduced density matrices with flat spectra (all nonzero eigenvalues are equal). Therefore, nonzero $\mathcal{AF}$ is an indirect proxy of non-stabilizerness. However, unlike the 
stabilizer R\'enyi entropy $\mathcal{M}_2$ that we compute via the Pauli--Markov chain sampling, 
anti-flatness is a subsystem quantity: it depends on the choice of bipartition and is sensitive to boundary effects. This quantity has been studied in the context of gauge theories as well~\cite{Grieninger:2026bdq}. We use a gauge-invariant definition of anti-flatness like entanglement entropy. For the dressed $\mathrm{SU}(2)$ model with $\jmax = 1/2$, the boundary cut can carry at most two terms. In the strong-coupling limit ($g \to \infty$), the ground state approaches a product state with $p_0 = 1$, $p_{1/2} = 0$, so $\mathcal{AF} \to 0$.
Similarly to the entanglement entropy computation, the gauge-invariant anti-flatness should be computed sector by sector as: 
\begin{equation}
    \mathcal{AF} = \sum_j p_j^2 \, \mathcal{AF}\left(\rho_A^{(j)}\right),
\end{equation}
or explicitly,
\begin{equation}
     \mathcal{AF} = \sum_j p_j^2
\left[
        \mathrm{Tr}\!\left(\rho_A^{(j)}\right)^3
        -
        \left(
            \mathrm{Tr}\!\left(\rho_A^{(j)}\right)^2
        \right)^2
    \right].
\end{equation}
This definition measures the non-flatness of the entanglement spectrum within each physical superselection sector and provides a lower bound to the non-local magic~\cite{Cao:2024nrx, Grieninger:2026bdq}. We show the results obtained in Fig.~\ref{fig:AF_and_its_ratio_with_SRE}. The ratio between the lower bound of non-local magic and SRE varies between $2\%$
and $24\%$ with a maximum at small $m$ and large $g$.

\end{document}